\documentclass{article}

\usepackage{PRIMEarxiv}
\usepackage{natbib}
\usepackage{array}
\usepackage{ragged2e,tabularx}
\usepackage{longtable}

\usepackage[utf8]{inputenc} 
\usepackage[T1]{fontenc}    
\usepackage{hyperref}       
\usepackage{url}            
\usepackage{booktabs}       
\usepackage{amsfonts}       
\usepackage{nicefrac}       
\usepackage{microtype}      
\usepackage{lipsum}
\usepackage{fancyhdr}       
\usepackage{graphicx}       
\graphicspath{{media/}}     

\title{A systematic review of early warning systems in finance
}

\author{
  Ali Namaki, Reza Eyvazloo \\
  Assistant Prof., Faculty of Finance and Insurance \\
  University of Tehran \\
  Tehran, Iran\\
  \texttt{AliNamaki@ut.ac.ir, Eivazlu@ut.ac.ir} \\
   \And
  Shahin Ramtinnia \\
  Ph.D. candidate, Faculty of Finance and Insurance \\
  University of Tehran \\
  Tehran, Iran\\
  \texttt{Shahin.Ramtinnia@ut.ac.ir} \\
}

\begin{document}
\maketitle

\begin{abstract}
Early warning systems (EWSs) are critical for forecasting and preventing economic and financial crises. EWSs are designed to provide early warning signs of financial troubles, allowing policymakers and market participants to intervene before a crisis expands. The 2008 financial crisis highlighted the importance of detecting financial distress early and taking preventive measures to mitigate its effects. In this bibliometric review, we look at the research and literature on EWSs in finance.

Our methodology included a comprehensive examination of academic databases and a stringent selection procedure, which resulted in the final selection of 616 articles published between 1976 and 2023. Our findings show that more than 90\% of the papers were published after 2006, indicating the growing importance of EWSs in financial research. According to our findings, recent research has shifted toward machine learning techniques, and EWSs are constantly evolving. We discovered that research in this area could be divided into four categories: bankruptcy prediction, banking crisis, currency crisis and emerging markets, and machine learning forecasting. Each cluster offers distinct insights into the approaches and methodologies used for EWSs.

To improve predictive accuracy, our review emphasizes the importance of incorporating both macroeconomic and microeconomic data into EWS models. To improve their predictive performance, we recommend more research into incorporating alternative data sources into EWS models, such as social media data, news sentiment analysis, and network analysis.
\end{abstract}

\keywords{Early Warning Systems(EWS) \and Finance \and Financial Crisis \and Economics \and Systematic Review}

\section{Introduction}

Financial early warning systems (EWS) are critical for spotting possible economic instability and reducing the likelihood of financial catastrophes \citep{Laeven}. The 2008 financial crisis highlighted the need for early financial distress identification and the significance of implementing preemptive actions to lessen its effects \citep{Demirguc,Mishkin}. EWSs are intended to offer early warning signs of financial troubles so policymakers and market players may take action before a crisis expands \citep{Claessens}. The literature on EWS in finance has dramatically changed over the last several decades as scholars have investigated different methodologies and approaches to creating EWS models that are more precise and efficient \citep{Kim}. 

Early research on EWS in finance was mainly concerned with creating econometric models to foretell financial disasters \citep{Bussiere1}. These models depended on macroeconomic data like interest and GDP growth rates to forecast financial instability. However, as methods for gathering and analyzing data have advanced, researchers have begun to focus on adding microeconomic data, such as the financial details of specific businesses, into the EWS models \citep{Berg}. Due to the broader perspective it offers of the economic environment, these methods have shown to be more successful in anticipating financial crises \citep{Drehmann}. 

Another trend in the literature is the increased emphasis on the role of financial interconnections and network effects in the transmission of financial risk \citep{Gai}. Network-based EWS models, which consider the linkages between financial institutions and their potential influence on the transmission of financial risk, have been developed as a result of this understanding \citep{Battiston1}. Researchers have recently been able to create more complex EWS models that combine network-based analysis, machine learning algorithms, and big data analytics because of advancements in data gathering and analysis methodologies \citep{Bargigli,Wever}.

Due to their capacity to deliver early warnings of potential threats in various industries, early warning systems (EWSs) have recently attracted much attention. Systemic risk, credit risk management, stock market research, banking supervision, market risk management, insurance, public finance, environmental finance, and supply chain finance are just a few areas where EWSs have been used in the finance industry \citep{Anwar}. EWS models are used in credit risk management to track individual enterprises' creditworthiness and spot impending defaults \citep{Altman1}.EWS models have been applied to stock market analysis to find future investment opportunities and impending market crashes \citep{Alshater,Samitas}. EWSs are employed in banking supervision by regulatory agencies to track the state of the banking industry and spot potential systemic hazards \citep{Borio1}. The risk of the entire financial system collapsing due to links and dependencies among financial institutions and between financial institutions and the real economy is referred to as systemic risk \citep{Battiston1}. Providing early warnings and implementing preventative measures before the dangers become too great, EWSs can be extremely important in controlling systemic risk.

Additionally, EWSs are being utilized more frequently in surveillance and fraud detection to find outlandish trends or behaviors that might be signs of fraud \citep{Baesens2,Liang}. In the insurance sector, EWSs are utilized to monitor policyholder behavior and assess potential claims \citep{Ndaru}. while in microfinance, they are used to assess loan portfolio risks and identify potential defaults \citep{Pakidame}.

Similarly, in public finance, EWSs are employed to track the financial well-being of sovereign states and flag possible debt defaults \citep{Zhao}. Supply chain financing also utilizes EWSs to monitor the risk of supply chain disruptions and identify potential stability risks \citep{Yin}.

This bibliometric review aims to give a broad overview of state of the art for EWSs in the finance sector, covering applications, trends, and significant historical contributions. Using a bibliometric technique, we gathered 616 articles from journals indexed in the Scopus database published from 1976 through 2023. This study's remaining sections are as follows; The data and methods of the investigation are covered in Part 2, the research's results are evaluated in Section 3, and the conclusion is presented in Section 4.

\section{Data and Methodology}
\label{sec:headings}
\subsection{Data Collection}
We gathered our data from the Scopus database. We made use of this database for a variety of purposes. It is one of the best databases for locating the most reliable publications in the banking and finance industry. It offers users different facilities for bibliometric analysis, such as exporting bibliographical data based on users' requirements.  

\subsection{Query}
 Based on their relevance to our research questions, we chose the search terms "early warning system," "finance," "bank," "bankruptcy prediction," "stock market," "systemic risk," "fraud detection," and more relative terms. These terms reflect the key concepts and approaches in the financial early warning systems literature.

 We used the Scopus database to conduct our search, which is widely regarded as a reliable source of scholarly literature in the banking and finance industry. Using the search terms, we found 5,402 articles on Scopus. We then screened the articles and we chose 616 articles for review after the screening. Table \ref{Table 1} summarizes the screening process by showing the number of papers at each stage.

We recognize that our search strategy has limitations, such as the exclusion of articles published outside of Scopus and the possibility of missing relevant articles due to insufficient literature coverage. We believe, however, that our search strategy provides a thorough overview of the literature on early warning systems in finance. 

\begin{table}[]
    \centering
    \begin{tabular}{c >\justifying m{10cm} c}
    \hline
    \textbf{NO.} & \textbf{Search Query} & \textbf{Results}\\
    \#1 & \textbf{Final Query Based on related keywords:}
    (TITLE-ABS-KEY ("early warning") OR TITLE-ABS-KEY ("early warning system" ) OR TITLE-ABS-KEY ("ews")) AND (TITLE-ABS-KEY("finance") OR TITLE-ABS-KEY("financial")  OR  TITLE-ABS-KEY ("banking") OR TITLE-ABS-KEY ("bank") OR TITLE-ABS-KEY ("stock") OR TITLE-ABS-KEY("stock market") OR TITLE-ABS-KEY("systemic risk") OR TITLE-ABS-KEY("fraud detection") OR  TITLE-ABS-KEY ("compliance") OR TITLE-ABS-KEY("algorithmic trading") OR TITLE-ABS-KEY("algo trading") OR TITLE-ABS-KEY("surveillance") OR TITLE-ABS-KEY ("portfolio management" ) OR TITLE-ABS-KEY("market risk") OR TITLE-ABS-KEY("credit risk") OR TITLE-ABS-KEY("liquidity risk") OR TITLE-ABS-KEY("model risk") OR TITLE-ABS-KEY("risk aggregation") OR  TITLE-ABS-KEY("stress testing") OR TITLE-ABS-KEY("behavioral finance") OR TITLE-ABS-KEY("sentiment analysis"))  & 5402 \\
    \#2 & \textbf{Limited results in journals, articles, reviews and English:} AND (LIMIT-TO(SRCTYPE, “j")) AND (LIMIT-TO(DOCTYPE, "ar") OR LIMIT-TO(DOCTYPE, "re")) AND (LIMIT-TO(LANGUAGE, "English")) & 3491 \\
    \#3 & \textbf{Limited subject area to the business, management and accounting(busi) or economics, econometrics and finance(econ):} AND SUBJAREA (busi  OR  econ) & 680 \\
    \#4 & \textbf{Manual filteration} & 616 \\
    \hline
    \end{tabular}
    \caption{Query and refinement used for data collection}
    \label{Table 1}
\end{table}

Selected articles were analyzed by popular methods and approaches in the field of bibliometrics, such as RStudio (bibliometrix library and biblioshiny app) and VOSViewer.

\subsection{Study approach and tools}
The study of scientific literature and the application of quantitative techniques to assess the significance and visibility of scientific research are known collectively as bibliometrics or scientometrics. Bibliometric approaches have been widely used in various fields, including information science, library and information studies, and science and technology studies, to measure the impact and productivity of authors, institutions, journals, and countries. Bibliometrics is a potent tool for comprehending the past and present state of research and trends and patterns in creating and disseminating scientific knowledge.

The history of bibliometrics can be traced back to the 20th century, with the introduction of citation analysis as a tool for measuring the impact of scientific research. Pritchard introduced the bibliometric approach for the first time to recognize and comprehend the network of articles based on citations \citep{Pritchard}.

Vast amounts of data on scientific literature can now be analyzed thanks to the development of digital libraries and the widespread use of online databases, which has dramatically increased the potential for bibliometric analysis. New techniques and tools for bibliometric analysis have been created due to the growing use of bibliometrics in research assessment, including co-citation analysis, co-word analysis, and network analysis \citep{Moed}.

To improve the review process's transparency, rigor, and accuracy, there has been an increase in interest in using bibliometric approaches for systematic reviews and meta-analyses in recent years \citep{Bornmann}. The selection and screening of articles, the gathering of data and information, and the evaluation of the standard and applicability of the literature have all been supported by bibliometric approaches\citep{Pulsiri}.

To thoroughly review the literature on EWSs in finance, we used a bibliometric approach in this work. Our goal was to present an in-depth analysis of this area's current state of the art and pinpoint the significant research axes and trends. Following accepted bibliometric guidelines and practices, we used a strict and open methodology. Utilizing various tools and methods, such as network analysis, co-citation analysis, and citation analysis, was part of our bibliometric analysis. The findings of our study and a discussion of the critical conclusions and their ramifications for future research are presented in the following section. 

\section{Results and discussion}
\label{sec:others}
\subsection{General analysis}
The search was carried out in February 2023 using Scopus as the database for this review. As explained in Table \ref{Table 1}, our dataset included 616 documents from 294 sources (including 605 articles and 11 reviews), and the research we gathered was done between 1976 and 2023. Other critical general data is shown in Table \ref{Table 2}.

\begin{table}[]
    \centering
    \small
    \begin{tabular}{|c|c|c|c|}
    \hline
    Timespan & 1976:2023 & Average citation per doc & 19.26 \\
     \hline
    Sources & 294 & Authors & 1316 \\
     \hline
    Documents & 616 & Total author`s keywords & 1612 \\
     \hline
    The annual growth rate of published documents & 3.48\% & Document average age(year) & 8.32 \\
     \hline
    Authors of single-authored docs & 107 & International Co-Authorship & 24.51\% \\
     \hline
    Co-Authors per Doc & 2.5 & Author's Keywords & 1612 \\
     \hline
    \end{tabular}
    \caption{General information of investigated database}
    \label{Table 2}
\end{table}

\subsubsection{Literature growth}
In this section, we present a comprehensive analysis of the temporal trends in the publication of EWS-related research. Fig.1 depicts the annual total number of articles published, the reverse cumulative sum percentage, and the 5-year moving average. The highlighted years show major financial crises happened in this era (1998, 2001, 2008, and 2020), showing that significant increases in the number of scientific articles may tend to follow financial crises, implying a relationship between the occurrence of financial crises and the subsequent growth of research in EWSs.

\begin{figure}
    \centering
    \includegraphics{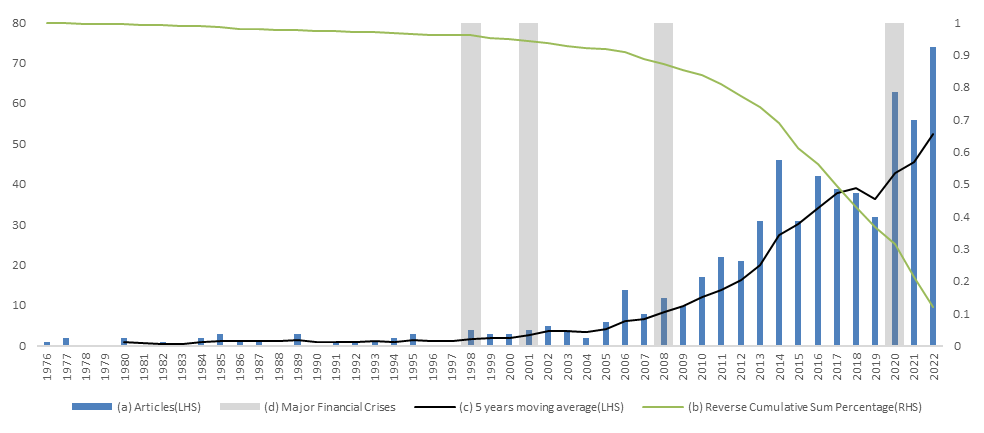}
    \caption{Temporal trends in EWS research publications from 1976 to 2023. The figure displays four key elements: (a) the total number of articles published per year, (b) the reverse cumulative sum percentage, (c) the 5-year moving average, and (d) highlighted years of major financial crises.}
    \label{Fig1}
\end{figure}

The Long-Term Capital Management (LTCM) hedge fund failed in 1998, which led to a review of risk management procedures in the financial industry \citep{Baesens1}. Due to the financial crisis, there is more interest in developing EWS for risk management now than ever.

The dot-com bubble burst in 2001, failing numerous technology companies and causing a precipitous drop in the stock market \citep{Stiglitz}. EWS for stock market analysis and the prediction was created in the wake of this crisis. The failure of several important financial institutions during the 2008 financial crisis was accompanied by a subsequent global recession. The crisis made it clear that the financial sector needed better risk management, and it stimulated more research on EWS for spotting systemic risk  \citep{Reinhart}. Finally, the recent COVID-19 pandemic in 2020 has created a global financial crisis, with many businesses and economies suffering significant losses \citep{Wahrungsfonds}. The pandemic has again brought to the forefront the importance of EWS in detecting potential threats to financial stability and, thus, increased research in this field.

Our EWS research publication trends analysis focused on the 2006-2022 period, which accounts for more than 90\% of total articles. Fig.2 depicts the number of articles published each year, the mean, and the one standard deviation threshold during this period. Notably, the number of articles published has increased over the years, with the one standard deviation threshold being exceeded in 2019-2022. This increase in publications and consistently exceeding the one standard deviation threshold during these years may demonstrate a growing interest in this research area.

\begin{figure}
    \centering
    \includegraphics[width=14cm]{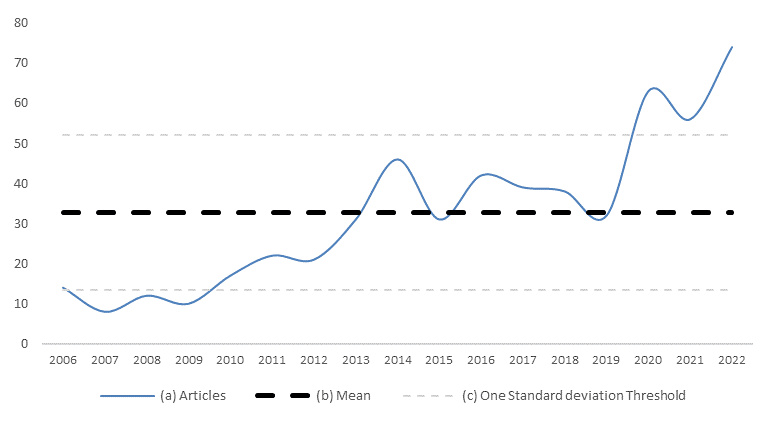}
    \caption{Publication trends in EWS research for the 2006-2022-time frame. The figure shows (a) the number of articles published per year, (b) the mean number of articles in this time frame, and (c) the one standard deviation threshold.}
    \label{Fig2}
\end{figure}

Several factors can be attributed to the increase in articles published on EWSs. As previously discussed, the increasing complexity of financial systems and the occurrence of major financial crises have prompted researchers to investigate new methodologies and data sources to improve the accuracy and reliability of EWSs. The increasing trend in publications emphasizes the importance of developing more robust and adaptive early warning systems to predict better, prevent, and mitigate future financial crises.

Second, the increasing data availability and advances in computational methods, particularly machine learning, have allowed researchers to investigate novel approaches to developing EWSs. Machine learning techniques can handle large amounts of complex data and have the potential to improve financial crisis prediction accuracy significantly \citep{varian}. This increased emphasis on machine learning in EWS research has contributed to increased publications between 2019 and 2022.

\subsubsection{An analysis of the sources}
This section presents an overview of the most influential sources in the EWS research field based on the number of articles published and total citations received. Fig.3 depicts these two metrics for the top 13 sources, revealing which journals have significantly contributed to EWS research.

According to our findings, the Journal of Financial Stability and the Journal of Banking and Finance are the most authoritative sources in terms of both the number of articles published and the total number of citations received. The Journal of Financial Stability is a well-known source of EWS research, with 23 articles and 1,056 citations. The Journal of Banking and Finance comes in second, with 22 articles and a total citation count of 1,469. These two journals are well-known for their emphasis on financial stability and banking, making them especially relevant for EWS research.

With 17 and 14 articles, respectively, the International Journal of Finance and Economics and the International Review of Financial Analysis contribute significantly to the EWS literature. However, their total citation counts of 310 and 246 indicate that their impact may be less than that of the leading sources. Nonetheless, these journals provide valuable perspectives on EWS, particularly from international finance and economics perspectives.

The Journal of International Money and Finance, with nine articles and 474 citations, and the International Journal of Forecasting, with eight articles and 213 citations, are two other notable sources. In the context of EWS research, these journals emphasize the importance of international money markets and financial forecasting.

It is worth noting that some publications, such as Applied Economics Letters and Applied Economics, have published a relatively higher number of articles (10 and 8, respectively) but have received a relatively lower number of citations (44 and 55, respectively). This finding could imply that, despite their contributions to the field, the influence of these sources on EWS research is less significant than that of the aforementioned leading journals.

\begin{figure}
    \centering
    \includegraphics{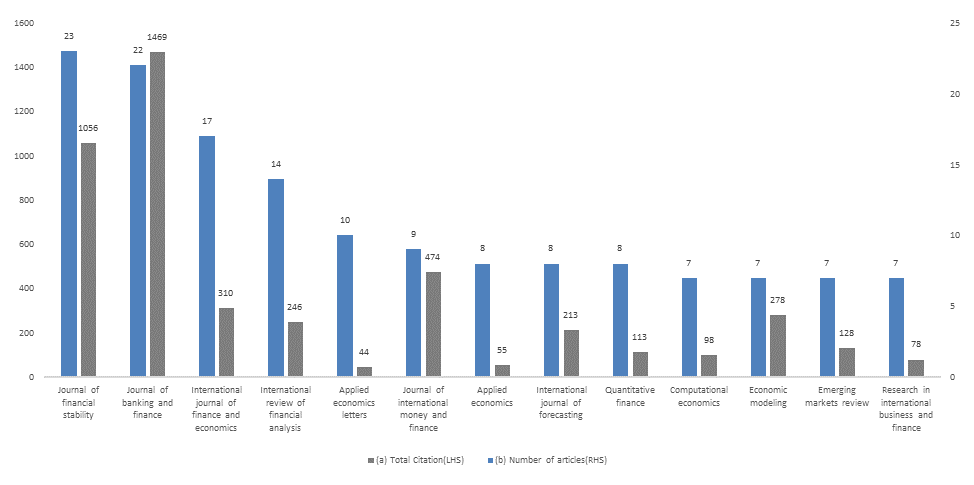}
    \caption{Top sources in EWS in finance research area based on the (a) number of articles published (RHS) and (b) total citations received (LHS)}
    \label{Fig2}
\end{figure}

\subsubsection{An analysis of the authors}
The analysis of authors, which was done with the help of the biblioshiny package, is presented in this part. The authors' analysis sheds light on the patterns of authorship and collaboration in the subject area. As shown in Fig. 4, The top writers in this field are represented by their total number of citations and h-index in the graph. Both metrics are local because they were computed using the articles and documents retrieved from our bibliometric search. The writers are presented in descending order of their h-index, which indicates their level of influence in the field.

\begin{figure}
    \centering
    \includegraphics{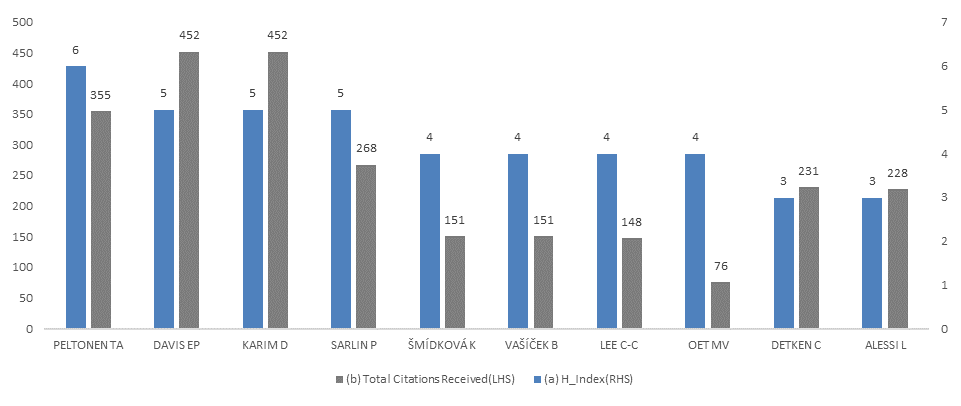}
    \caption{Top 10 authors in EWS in finance research area, based on the (a) H-index (RHS) and (b) total citations received (LHS) within the collection of articles from bibliometric search.}
    \label{Fig4}
\end{figure}

With an H-index of 6 and 355 citations, Peltonen, T.A. emerges as a leading author in EWS research, according to our analysis. The H-index is a well-known metric that measures a researcher's productivity and citation impact, indicating that Peltonen's research output is substantial and highly influential in the field.

Davis, E.P., and Karim, D. have the same H-index of 5 and a total citation count of 452, making them the following most influential authors in EWS research. Their contributions have been widely cited in the literature, demonstrating their significant influence on developing and understanding financial early warning systems.

P. Sarlin is another notable author in the field, with an H-index of 5 and 268 citations. Sarlin's work has been influential and well-recognized within the EWS research community, despite having a lower total citation count than Davis and Karim.

Other authors who have made significant contributions to the field of EWS research include mdková, K., and Vaek, B., both of whom have an H-index of 4 and 151 citations, and Lee, C-C., who has an H-index of 4 and 148 citations. As evidenced by their citation metrics, these authors have made significant contributions to the literature.

Detken, C., and Alessi, L., on the other hand, have a lower H-index of 3 but have received more total citations (231 and 228, respectively). While their publication output may not be as extensive as other authors on the list, their work has significantly impacted the field.

In conclusion, our analysis of the authors highlights the key researchers who have made significant contributions to EWS research and their respective impact on the field. Recognizing these influential authors can assist researchers in identifying and accessing the most relevant and impactful literature on EWSs in finance and recognizing these individuals' contributions to advancing the understanding of early warning systems in the finance domain.

Fig.5 depicts the authors' output over time and highlights an exciting trend in the timeline of articles that have been published. It is clear that starting in 2006 and onward, the EWSs in the financial sectors received the majority of attention. This may be connected to using artificial intelligence (AI) and machine learning in early warning systems (EWSs) and the gradual increase in computing power availability. Recent developments in AI and machine learning techniques may have allowed researchers to analyze large datasets and create more precise and efficient EWSs \citep{Hassanien}.

\begin{figure}
    \centering
    \includegraphics{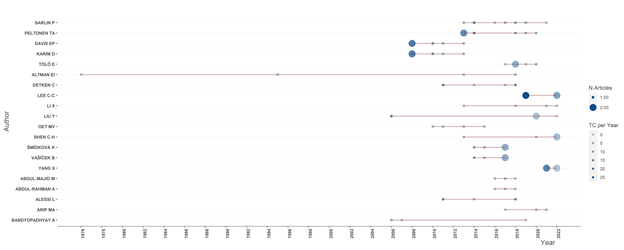}
    \caption{Timeline of authors' output in EWS in finance research from 1976 to 2023, with hue indicating the total citations per year and the size of the circles representing the number of articles published.}
    \label{Fig5}
\end{figure}

Additionally, the pattern in Fig. 5 might indicate a rise in research interest in EWSs. The fact that more articles have been published on EWSs since 2006 lends credence to this theory. Due to their widespread practical applications in industries like finance and healthcare, there is an expanding body of research on EWSs. The growth in authors contributing to the field is attributable to the growing number of authors with published works in recent years and the growing interest in EWSs as a research topic.

\subsubsection{An analysis of the countries}
In this section, we present an overview of the countries that have significantly contributed to the EWS in the finance research field based on the number of single-country articles, multiple-country articles, and total citations received within our bibliometric collection of articles. Figure 6 depicts these three metrics for the top ten countries, providing insights into the geographical distribution of EWS research and its impact on the field.

\begin{figure}
    \centering
    \includegraphics{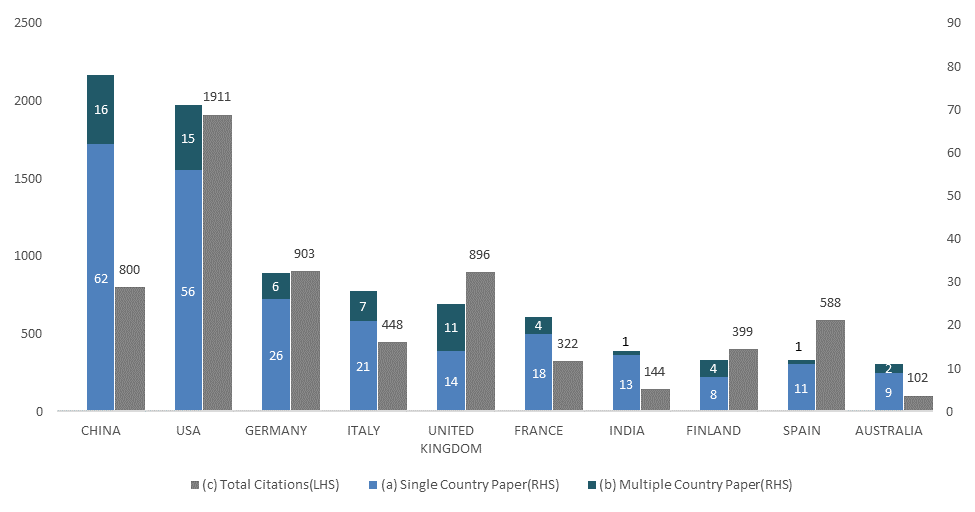}
    \caption{Top 10 countries in EWS research based on the number of (a) single country articles (RHS), (b) multiple country articles (RHS), and (c) total citations received (LHS)}
    \label{Fig6}
\end{figure}

According to our analysis (Fig. 6), China has the most single-country articles, with 62, closely followed by the United States, which has 56. However, the United States outnumbers China's total citations, with 1,911 citations to China's 800. This finding implies that, while both countries have made significant contributions to this research area, the impact of the United States' work appears more substantial.

Germany and Italy also significantly contribute to EWS research, with 26 and 21 single-country articles, respectively. Germany has 903 total citations, while Italy has 448. These countries have significant involvement in the field, with Germany's research output being incredibly influential based on citation count.

In summary, our country analysis highlights the geographical distribution of EWS research and the impact of the work done by researchers in these countries. The United States and China lead in publication output, while the United States, Germany, and the United Kingdom have significant influence based on citation counts. Recognizing these countries' contributions can assist researchers in identifying and accessing the most relevant and impactful literature on EWSs in finance and understanding the geographical distribution of research in the field.

\subsubsection{Most relevant affiliations }
According to the data of most relevant affiliations, exhibited in Table.3, a diverse set of affiliations contributes to the field, with some universities and institutions standing out as key players. Regarding the number of published articles, the University of California emerges as the leader, closely followed by Northeastern University and The Bucharest University of Economic Studies. These leading institutions emphasize the global nature of the research, with contributions from all over the world.

The presence of a central bank, the Deutsche Bundesbank, among the top contributors is an intriguing finding. This suggests that research in this area is essential for academic institutions, policymakers, and financial institutions, indicating the subject's broader scope and relevance.

Furthermore, the data shows that the number of articles published by the top institutions is relatively close. This indicates a competitive and collaborative research environment, with multiple organizations actively contributing to field knowledge advancement. Overall, the data emphasize the research's interdisciplinary and global nature, emphasizing the importance of collaboration between academic institutions and other organizations in driving progress.

\begin{table}[]
    \centering
    \small
    \begin{tabular}{|c|c|}
    \hline
     Affiliation & Number of articles  \\
     \hline
     University of California &  8\\
     \hline
     Northeastern university &  7\\
     \hline
     The Bucharest university of economic studies &  7\\
     \hline
     University Kebangsaan &  6\\
     \hline
     Brunel university &  5\\
     \hline
     Central university of finance and economics &  5\\
     \hline
     Deutsche bundesbank &  5\\
     \hline
     Fordham university &  5\\
     \hline
     Universidade nova de Lisboa &  5\\
      \hline
     Not reported &  5\\
     \hline
    \end{tabular}
    \caption{Most relevant affiliations}
    \label{Table 3}
\end{table}

\subsection{Citation analysis}
The top 10 documents in our bibliometric search are shown in table.\ref{Table 4} The topics covered in the papers are diverse and include banking supervision and systemic risk, financial crises, credit risk, stress testing, and financial market forecasting and prediction. These papers' high citation counts indicate that they have made significant advances in their respective fields of study, are regarded as seminal works, and have a lasting impact on subsequent research.

\begin{longtable}{>{\raggedright\arraybackslash}p{4cm}>{\raggedright\arraybackslash}p{4cm}>{\raggedright\arraybackslash}p{4cm}>{\raggedright\arraybackslash}p{2cm}}
\caption{Most cited documents}
\label{Table 4} \\
\toprule
\textbf{Title} & \textbf{Author(s)/Year} & \textbf{Source Title} & \textbf{Global Total Citations} \\
\midrule
\endhead

\endfoot

\bottomrule
\endlastfoot

{\footnotesize Early warning of bank failure: A logit regression approach} & \citet{Martin} & Journal of Banking and Finance & 1716 \\
{\footnotesize SRISK: A conditional capital shortfall measure of systemic risk} & \citet{Brownless} & The Review of Financial Studies & 1183 \\
{\footnotesize Credit risk in two institutional regimes: Spanish commercial and savings banks} & \citet{Salas} & Journal of Financial Services Research & 1353 \\
{\footnotesize Towards a new early warning system of financial crises} & \citet{Bussiere1} & Journal of International Money and Finance & 1017 \\
{\footnotesize Dating the timeline of financial bubbles during the subprime crisis} & \citet{Phillips} & Quantitative Economics & 659 \\
{\footnotesize An index of financial stress for Canada} & \citet{Illing} & Bank of Canada & 324 \\
{\footnotesize Can leading indicators assess country vulnerability? Evidence from the 2008–09 global financial crisis} & \citet{Frankel} & Journal of International Economics & 589 \\
{\footnotesize Comparing early warning systems for banking crises} & \citet{Davis} & Journal of Financial stability & 660 \\
{\footnotesize Market transparency and the accounting regime} & \citet{Bleck} & Journal of accounting research & 443 \\
{\footnotesize ‘Real time’ early warning indicators for costly asset price boom/bust cycles: a role for global liquidity.} & \citet{Alessi} & European Central Bank & 349 \\

\end{longtable}

From examining the top 10 most cited papers, it is clear that most have to do with banking supervision and systemic risk. The high citation count of these papers leads to the conclusion that the banking supervision and systemic risk area is an important research area that has attracted significant attention recently.

\subsection{Network visualization}
\subsubsection{Word cloud}
As shown in Fig.7, the most frequently occurring terms in the collection of articles found by the bibliometric search were used to create a word cloud for this study. The generated word cloud gives a quick overview of early warning systems literature's most famous subjects and ideas. 

\begin{figure}
    \centering
    \includegraphics{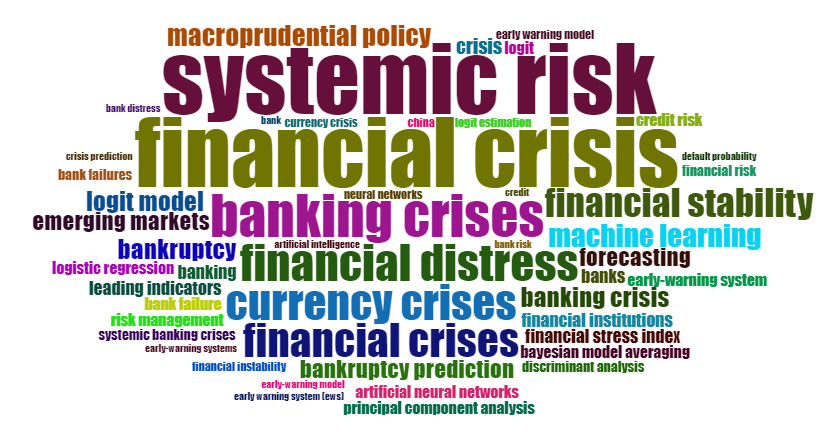}
    \caption{Word cloud}
    \label{Fig7}
\end{figure}

 The created word cloud (Fig.7) summarizes the key terms taken from the retrieved papers' titles and abstracts. The most popular search terms are "systemic risk", "financial crisis", "banking crisis", "financial distress", "machine learning models", and "currency crisis", as indicated in the figure. These terms highlight the primary area of interest in studying early warning systems and the value of researching the early warning and detection of financial crises and systemic risks. The word cloud's high frequency of phrases linked to numerical and machine learning models highlights the growing interest in using computational techniques and algorithms to increase early warning systems' accuracy.

\subsubsection{Trend topics}
The development of the most frequently discussed subjects in the examined documents is shown in Fig.8 The size of the circles in this diagram indicates how frequently the term occurs, and the length of the lines indicates how long it has been under study (by the first and third quantiles). It is drawn by using authors' keywords in the biblioshiny application by setting parameters such as word minimum frequency to 5 and number of words per year to 3.  

\begin{figure}
    \centering
    \includegraphics[width=14cm]{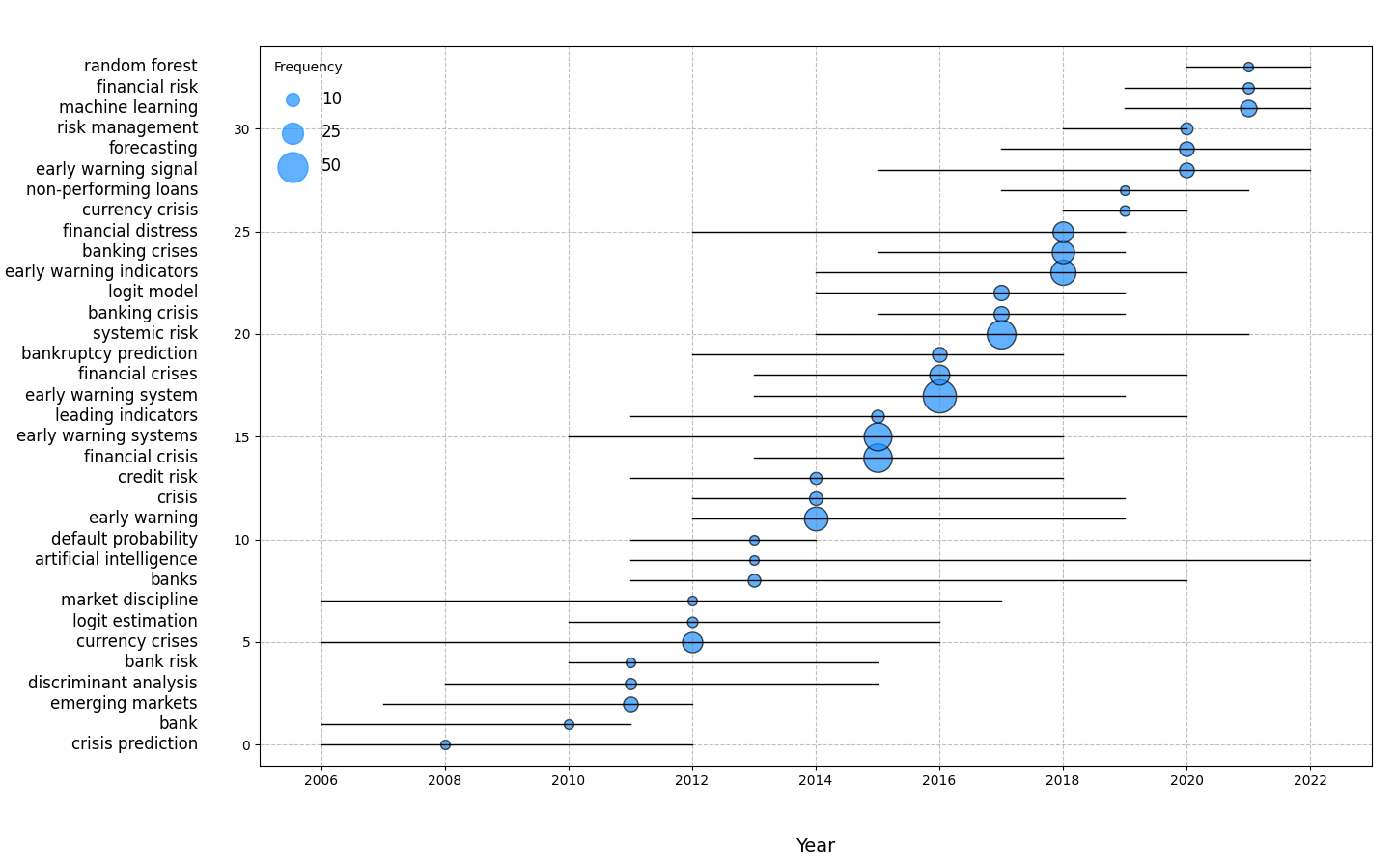}
    \caption{Trend topics}
    \label{Fig8}
\end{figure}

According to the trend topics analysis, machine learning, random forest, forecasting, and artificial intelligence are some of the most studied and discussed topics in the area of early warning systems for financial crises. This reflects a growing interest in using cutting-edge statistical and computational methods to create EWSs that are more precise and trustworthy.

Due to its capacity to learn from historical data and make predictions based on intricate patterns and relationships, machine learning, in particular, has attracted much attention recently. Because it can handle large amounts of data and non-linear relationships, the random forest machine learning algorithm has been extensively used in EWSs. Another crucial subject is forecasting, which is crucial for foreseeing and averting future financial crises. It involves estimating future values of critical economic indicators like GDP, inflation, and interest rates using statistical and econometric models.

Overall, the trend topics analysis indicates that as researchers and practitioners work to create more precise and reliable methods for anticipating and preventing financial crises, using advanced statistical and computational techniques is becoming increasingly significant in EWSs.

\subsubsection{Thematic analysis}
We use a thematic map in this section to gain insights into the main research themes and their relationships in the field of early warning systems (EWSs) in finance. A thematic map is a visualization tool that helps researchers identify and visualize the main research themes using keyword co-occurrence analysis \citep{Cobo}. A thematic map can provide an understanding of the structure of a research field and reveal trends, gaps, and potential areas for future research by examining the relationships between keywords in the scientific literature.

Centrality and density are two key metrics used in creating a thematic map. The importance of a theme in the research field, as determined by its connections to other themes, is referred to as centrality \citep{Bastian}. A higher centrality score denotes a more central and connected theme within the research field. On the other hand, density represents the strength of relationships between keywords within a theme, indicating the theme's coherence and internal development \citep{Cobo}. These two measures are used to create a thematic map by the R package "bibliometrix" with the following formulae:

$$ Centrality(C) = 10 \times \sum{e_{kh}}$$

$$ Density(D) = 100\sum{\frac{e_{ij}}{w}}$$

With k a keyword belonging to the theme and h a keyword belonging to other themes and i,j keywords belonging to the theme and w the number of keywords in the theme.

Degree centrality is the number of links a node (keyword) has to other nodes, whereas weighted degree is the sum of all link weights\citep{Aria}. A thematic map is a two-dimensional space where themes are plotted based on their centrality and density rank values (when using median for classification) or actual values (when using mean) along the x-axis (centrality) and y-axis (density). Four types of themes can be identified based on their position in the quadrants of the strategic diagram:

\begin{itemize}
  \item Upper-right quadrant (motor themes): These themes are both well-developed and important for the structure of the research field, known as motor-themes. They exhibit strong centrality and high density, indicating that they are related externally to concepts applicable to other closely related themes.
  \item Upper-left quadrant (niche themes): Themes in this quadrant have well-developed internal ties but unimportant external ties, making them marginal to the field. These themes are highly specialized and peripheral in character.
  \item Lower-left quadrant (emerging or declining themes): Themes in this quadrant are weakly developed and marginal, displaying low density and low centrality. They typically represent emerging or disappearing themes in the research field.
  \item Lower-right quadrant (basic themes): Themes in this quadrant have strong external ties (high centrality) but weak internal ties (low density). These themes are known as basic themes, which act as connectors between the motor themes and form the fundamental structure of the research field. They might not be well-developed themselves, but they play a crucial role in linking other themes and facilitating interdisciplinary connections.
\end{itemize}

In this paper, a thematic map was created to analyze the authors` keywords in the literature pertinent to our research topic using biblioshiny with 500 words (biblioshiny`s keyword plus) and four as minimum cluster frequency (per thousand docs) as shown in Fig.9. 

\begin{figure}
    \centering
    \includegraphics[width=14cm]{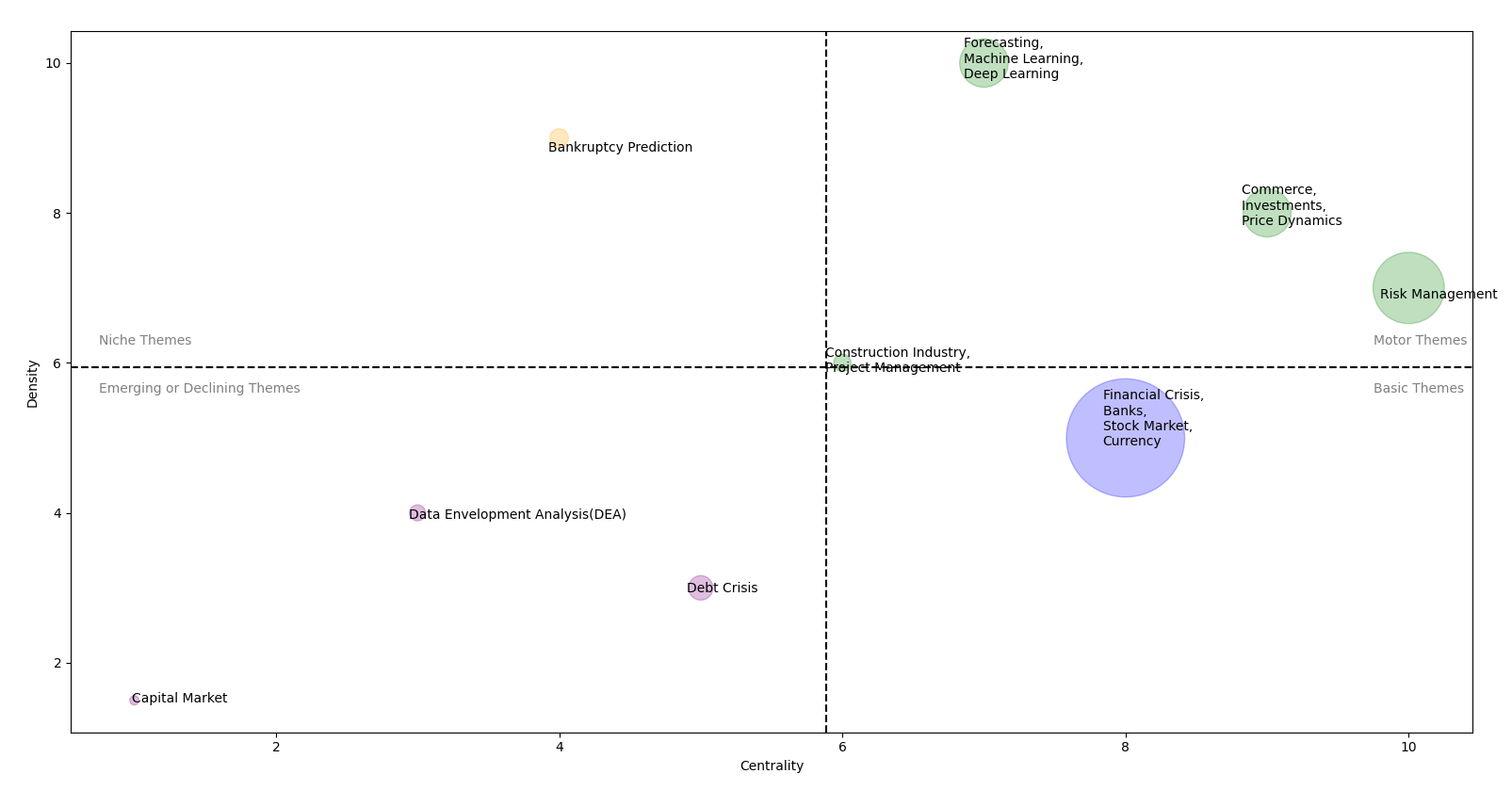}
    \caption{Thematic map, illustrating the research landscape in the EWS in finance research sector.}
    \label{Fig9}
\end{figure}

Several clusters have been identified due to the thematic map analysis, which provides insights into the research field. Three distinct clusters have emerged within the quadrant of motor themes:
\begin{itemize}
  \item \textbf{Forecasting, machine learning, and deep Learning:} This cluster emphasizes artificial intelligence and advanced computational techniques' growing role in the financial sector. These themes' prominence suggests they are well-developed and central to the field, driving innovations and advancements in financial forecasting, risk management, and other related areas.
  \item \textbf{Commerce, investment, and price dynamics:} This cluster represents the fundamental financial concepts that shape the research landscape, such as commerce, investment, and price fluctuations. These concepts are critical for comprehending market behavior, investment strategies, and price movement factors.
  \item \textbf{Risk management:} This cluster emphasizes the significance of risk management in the financial sector. As a driving theme, it demonstrates that risk management is an essential aspect of financial research, with ongoing developments and innovations in risk assessment, mitigation, and monitoring.
\end{itemize}

One cluster is present in the basic themes quadrant:
\begin{itemize}
  \item \textbf{Financial crisis, banks, stock market, currency:} This cluster connects motor themes by focusing on the broader financial landscape, including the effects of financial crises, the role of banks, stock market fluctuations, and currency dynamics. While less well-developed than motor themes, these topics are critical to comprehending the overall structure and relationships within the research field.

\end{itemize}

Three clusters have been identified within the quadrant of emerging or declining themes:
\begin{itemize}
  \item \textbf{Debt crisis:}This cluster indicates an increase or decrease in interest in studying debt crises, which could be influenced by recent economic events or shifts in research focus.
  \item \textbf{Data envelopment analysis (DEA):} This cluster indicates that DEA, a technique for measuring performance, is losing traction in the research field. This could be due to the rise of alternative methods or a shift in financial research priorities.
  \item \textbf{Capital market:} This cluster represents capital market interest, which may be an emerging or declining theme depending on the evolving research landscape and economic developments.
\end{itemize}
Finally, one cluster has been discovered in the niche themes quadrant:
begin{itemize}
\begin{itemize}
  \item \textbf{Bankruptcy prediction:} This cluster represents a subfield of study that focuses on forecasting bankruptcies. As a niche theme, this topic may have a small audience or scope within the larger research field.
\end{itemize}

\subsubsection{Keywords analysis}
The Keywords analysis sheds light on the most popular keywords and their connections in the literature on the subject of interest. This technique entails visualizing the co-occurrence of keywords in the chosen papers and spotting groups of related terms that can point to underlying themes and ideas. The articles that are gaining the most significant attention in the literature and the research landscape can both be better understood by examining the keywords that appear the most frequently. Researchers can learn more about how various topics are connected and the developing study fields by finding the linkages between keywords.

\begin{figure}
    \centering
    \includegraphics[width=14cm]{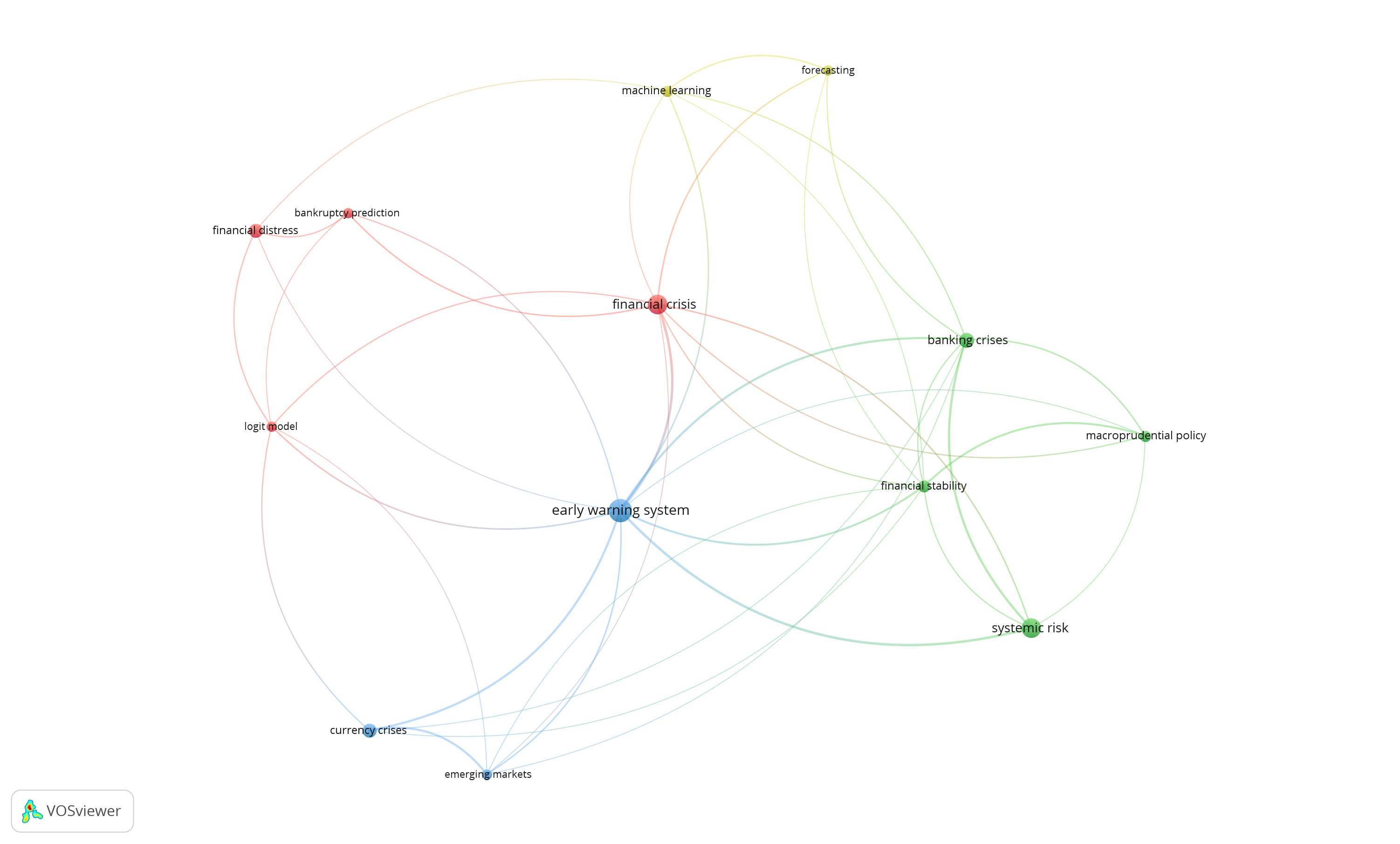}
    \caption{Keywords co-occurrence}
    \label{Fig10}
\end{figure}

As shown in Fig.10, The first cluster(red) includes the terms "bankruptcy prediction," "financial crisis," "financial distress," and "logit model." These terms suggest a preoccupation with foreseeing and averting financial crises, perhaps through creating models and techniques that might spot early warning indications of distress in financial institutions.

The second cluster(green), which also contains the terms "banking crisis," "financial stability," "macroprudential policy," and "systemic risk," points to a concentration on risk management in finance but with a stronger emphasis on regulatory frameworks and policy initiatives. The terms in this cluster emphasize systemic risk, the possibility that the entire financial system will collapse, and the requirement for macroprudential policies to maintain financial stability.

The third cluster(blue), including "currency crisis" and "emerging markets," emphasizes global finance and developing nations. The terms in this cluster point to an interest in exchange rates, financial instability, economic growth in emerging nations, and the possibility of currency crises in these settings.

The fourth cluster(yellow) suggests the concentration on quantitative techniques and data analysis, which consists of predicting and machine learning. These keywords suggest an interest in creating forecasting and prediction models and methodologies, maybe utilizing learning algorithms and other cutting-edge technologies.

Overall, the keyword co-occurrence chart indicates that risk management, financial crisis avoidance, and the development of quantitative tools for analysis and prediction are the main topics of interest in the literature on finance. 

The keyword co-occurrence chart's inclusion of a time dimension adds crucial details about the temporal development of the clusters as seen in fig10. The graph demonstrates that, compared to the other clusters, the one containing keywords about currency crises and emerging markets emerged earlier, around 2012. This suggests that, in contrast to the other clusters, these topics were of interest to researchers and industry professionals earlier. Around 2015, keywords with bankruptcy prediction and financial distress emerged. Around 2017, a keyword cluster containing terms associated with the banking crisis, financial stability, macroprudential policy, and systemic risk emerged. This cluster reflects the growing concern about the stability of the financial system. The most recent cluster to form around 2020 includes keywords associated with forecasting and machine learning, demonstrating the growing interest in using sophisticated analytical tools to analyze financial data.

\begin{figure}
    \centering
    \includegraphics[width=14cm]{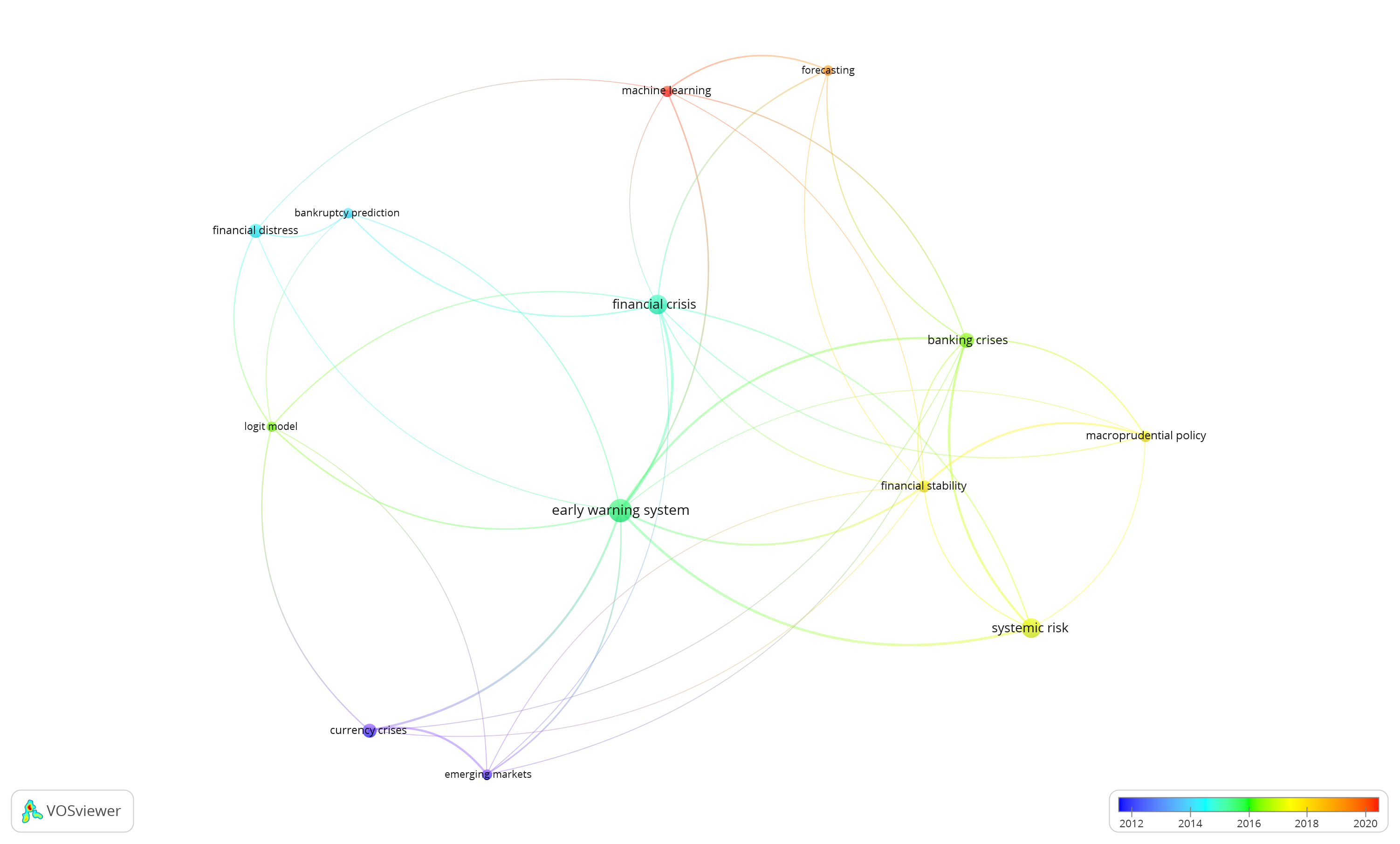}
    \caption{Overlay visualization}
    \label{Fig11}
\end{figure}

\subsection{Content analysis}

In this section, we present a content analysis of the primary research streams identified through our bibliometric analysis. As shown in Fig.12, we identified four major research streams within the financial sector by examining the keyword co-occurrence analysis. These research streams represent distinct but interconnected areas of study, demonstrating the breadth and depth of research in this field. We will overview each research stream and discuss the key papers and milestones that have shaped their evolution.

\begin{figure}
    \centering
    \includegraphics[width=14cm]{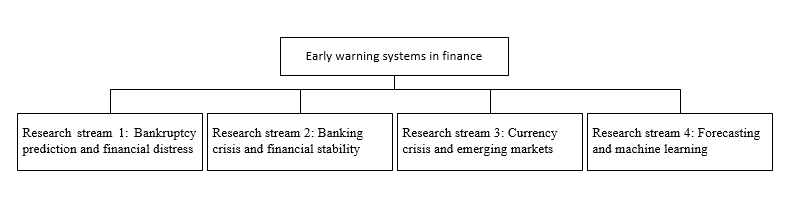}
    \caption{Research streams found in EWS in finance research area}
    \label{Fig12}
\end{figure}

\subsubsection{Research stream 1: Bankruptcy prediction and financial distress}
Predicting bankruptcy and financial distress is one of the critical study areas in the literature on early warning systems in finance. This research stream focuses on creating models and techniques to recognize businesses in financial crisis or danger of bankruptcy before these events occur.

The first study period in this research stream was primarily concerned with bankruptcy forecasting, and several significant articles emerged during this time. \citet{Altman1}, one of this field's oldest and most highly referenced works, introduced the Z-score model. This model, which has been used in various scenarios throughout the years, predicts company bankruptcy using financial ratios and discriminant analysis.

\citet{Ohlson}, which offered a multivariate bankruptcy prediction model that included accounting and market-based variables, was another important study from this era. The Ohlson model has undergone various revisions and additions since it first gained popularity as a method for predicting bankruptcy.

Researchers started investigating the use of artificial neural networks (ANNs) for bankruptcy prediction in the early 1990s, which resulted in the publication of several critical papers. Traditional statistical methods were outperformed by the model proposed by \citet{Odom}, which used neural networks to forecast bankruptcy.

Subsequent studies further confirmed the usefulness of neural networks in predicting business failure. \citep{Altman2,Fletcher,Lee,Tam2,Wilson}Besides ANNs, other machine learning and artificial intelligence-based methods have been adopted by many authors in the bankruptcy prediction area. Rough set theory, initially developed to deal with apparent indiscernibility between objects in a set, has been used in this area . \citep{Beynon1,McKee,Xiao}

Case-based reasoning has also become a crucial paradigm in business failure prediction due to its simplicity, competitive performance with newer methodologies, and ease of pattern maintenance. \citep{Li}. Lastly, support vector machines have outperformed artificial neural networks\citep{Kim2,Lin}. The numerous approaches utilized in this research stream show the ongoing search for reliable techniques for spotting financial trouble and foreseeing possible corporate bankruptcy.

\subsubsection{Research stream 2: Banking crisis and financial stability}
The global financial crisis of 2007–2008 made clear the necessity for an efficient Early Warning System (EWS) to prevent the risk of bank failures and runs from developing into systemic banking crises. The identification, justification, and prediction of crises have been the subject of a long tradition of writing. Systemic financial crises have gained attention during the past 20 years, replacing balance of payments and currency crises. However, economists offer a variety of criteria to pinpoint the incidence of financial crises because their definition is not simple.

The literature offers methods for establishing causal relationships between macroeconomic imbalances and crisis episodes, examines their function as leading indicators, and evaluates the capacity of econometric models to forecast banking crises or the likelihood of a crisis. Building a variable that identifies these episodes is helpful for the "estimation" of empirical models, or EWSs, intended to find the likelihood that a systemic banking crisis will occur, in addition to tracking the occurrence of such episodes.

The literature on EWSs has developed in different directions, starting with the signals approach and moving on to discrete choice models and machine learning methods. The signals approach is a non-parametric technique that examines the ex-post behavior of macroeconomic variables and checks to see if the indicators exhibit a pattern before crises that is distinct from one during quiet or standard times. If a variable rise above a certain level, it indicates a crisis. The model does not permit the aggregation of the individual warnings, and with the signals approach, each indicator is used singly.

\citet{Kaminsky3} first applied the signals approach to 76 currency crises and 26 banking crises between 1970 and 1995 in various industrial and developing countries.

The forecasting accuracy of each variable is considered by \citet{Kaminsky2}, among others, by creating a composite index that weights the signals of each variable by the inverse of their noise-to-signal ratios. This methodology is used to study banking crises from 1960–1999 and 1979–2003, respectively.\citep{Borio2,Davis}

The logit model, a tool frequently used in microeconometrics to estimate the probability of an event, is an alternative methodology that enables the simultaneous study of macroeconomic variables as determinants of banking crises. The binary outcome variable (crisis/non-crisis) is estimated as a function of macroeconomic factors, and the likelihood that the event (crisis) occurs is also estimated. The estimated crisis probabilities can be retrieved from the model's estimated coefficients.

Applying this methodology to a sizable sample of developed and developing nations between 1980 and 1994, \citet{Demirguc} find that low growth, high inflation, and high real interest rates are the primary causes of a banking crisis.

With contributions for a variety of nations and periods, the literature developed along these lines. \citep{Barrell,Demirguc2,Eichengreen,Schularick}

\subsubsection{Research stream 3: Currency crisis and emerging markets}

Early research on emerging markets mainly concentrated on currency crises and their effects. Investigating the causes of currency crises in developing nations by \citet{Kaminsky} is one of this field's earliest and most important works. The authors identified several indications, such as high levels of external debt and low reserves, that forecast the possibility of a currency crisis. Future studies on currency crises in emerging markets will be built on the findings of this study.

Early in the new millennium, experts examined how currency crises affected financial institutions and the economy. The research conducted by \citet{Eichengreen2} on banking crises in emerging nations made a significant addition to this field. The authors stated that a range of mechanisms, including capital flight and a decline in asset values, can cause banking crises brought on by currency crises.

Scholars have more recently looked into how macroeconomic policies might help emerging nations avoid and manage currency crises. \citet{Jeanne} investigation of the efficiency of macroprudential measures in lowering systemic risk is a landmark study in this field. The authors discovered that loan-to-value ratios and reserve requirements could help lower the likelihood and severity of currency crises. This line of research has significant ramifications for policymakers in emerging nations attempting to maintain financial stability in the face of external shocks.

\subsubsection{Research stream 4: Forecasting and machine learning}
Our bibliometric review's fourth research stream focuses on using machine learning methods to predict financial crises and advance early warning systems (EWSs). In recent years, this field of study has experienced significant growth, thanks to machine learning algorithms' strong potential to improve forecasting accuracy and offer insightful analyses of intricate financial systems.

\citet{Kaminsky} carried out one of the early studies in this field and were the ones to apply machine learning in the context of forecasting currency crises. Since then, a growing body of literature has examined different machine learning techniques and their applicability for EWSs in finance.\citep{Hillegeist,Liu}

Support vector machines (SVM), artificial neural networks (ANN), decision trees, and ensemble approaches are just a few machine learning techniques used on EWSs recently. SVM was employed by \citet{Huang} to forecast financial hardship in Taiwan's commercial banks with greater accuracy than conventional statistical models. \citet{Alfaro} used ensemble approaches to predict financial hardship, demonstrating the advantages of mixing many models to increase accuracy. 

Artificial neural networks have also been used in several EWSs, as shown by \citet{Boyacioglu}, who created an ANN-based model for foretelling financial crises in developing economies.
When compared to traditional linear regression models, their model performed better. To further advance machine learning in EWSs, \citet{Wang} utilized deep learning methods to predict bankruptcy.

\section{Conclusion}
Our bibliometric review revealed the significance of early warning systems in predicting and preventing financial crises. However, several obstacles must be overcome to improve the effectiveness and dependability of these systems. Utilizing machine learning techniques in creating EWSs is essential for future research. Our analysis reveals that machine learning algorithms, such as decision trees, neural networks, and support vector machines, have great potential for enhancing the precision and efficacy of EWSs. Therefore, additional research should be conducted to develop and evaluate EWS models based on machine learning in various financial contexts.

Creating EWSs for banking crises and financial stability is an additional important area for future research. Several financial crises in recent decades revealed significant weaknesses in the financial sector. Developing effective EWSs for identifying and preventing banking crises is therefore essential. In developing EWSs for banking crises, conventional financial ratios and macroeconomic indicators are frequently employed, according to a review of the relevant literature. Recent research indicates, however, that incorporating non-financial information, such as governance structures and risk management practices, can improve the accuracy and efficacy of EWSs.

Moreover, our analysis reveals the need for additional research on EWSs for currency crises and emerging markets. Emerging markets are frequently characterized by high volatility, limited data availability, and political instability, making the development of trustworthy EWSs complex. Future research should therefore concentrate on developing new models and techniques that can capture the unique characteristics of emerging markets and provide early warning signals for currency crises.

Incorporating social media data, news sentiment, and network analysis is another area that has gained traction in recent years for EWS. Social media platforms and news organizations provide much real-time data that can provide insights into market sentiment and potential risks. By analyzing the interconnections between financial institutions and the financial system as a whole, network analysis can assist in identifying systemic risk. However, while these data sources have shown promise, incorporating them effectively into EWS models requires more work. Important factors that must be addressed to ensure the accuracy and dependability of these models include data quality, bias, and the need for sophisticated data processing techniques. Therefore, additional research is required to develop robust EWS models that effectively integrate these data sources.

Finally, future research on EWSs could address several challenges, including the problem of data availability and quality, the need for additional testing and validation, and the difficulty of incorporating qualitative aspects into the models. These obstacles can reduce the efficacy and dependability of EWSs; therefore, they must be addressed to improve their predictive performance of EWSs.

In conclusion, our literature review demonstrates the significance of EWSs in predicting and preventing financial crises. However, some obstacles must be overcome to improve the effectiveness and dependability of these systems. Future research should therefore concentrate on developing and evaluating new models and techniques, addressing data availability and quality concerns, and incorporating qualitative aspects into the models. 

\bibliographystyle{plainnat}
\bibliography{references.bib}

\end{document}